\documentclass[showpacs,twocolumn,10pt,prl]{revtex4}
\usepackage{amsmath,graphicx}
\usepackage{amssymb}
\usepackage{pdfsync}
\def\ff{\textbf{f}}
\def\mm{\textbf{m}}
\def\pp{\textbf{p}}

\def\rr{\textbf{r}}
\def\BB{\textbf{B}}

\def\EE{\textbf{E}}
\def\HH{\textbf{H}}
\def\ss{\boldsymbol{\sigma}}
\def\s{\sigma}
\def\bra{\langle}
\def\ket{\rangle}
\def\f{\hat{f}}

\newcommand{\pder}[2]{{\frac{\partial #1}{\partial #2}}}

\begin{document}

\title{Quantum Kinetic Theory of Current-Induced Torques in Rashba Ferromagnets}

\author{D. A. Pesin and A. H. MacDonald }
\affiliation{Department of Physics, The University of Texas at Austin,  Austin TX 78712 USA}

\begin{abstract}
Motivated by recent experimental studies of thin-film devices containing a single ferromagnetic layer,
we develop a quantum kinetic theory of current-induced magnetic torques in Rashba-model
ferromagnets.  We find that current-induced spin-densities that are responsible for the switching behavior are due most essentially to spin-dependent quasiparticle lifetimes and derive analytic expressions for relevant limits of a simple model.  Quantitative model parameter estimates suggest that spin-orbit coupling in the adjacent metal normal magnetic layer plays an essential role in the strength of the switching effect.
\end{abstract}

\date{\today}
\maketitle

\begin{figure}
\begin{center}
\includegraphics[width=3.0in]{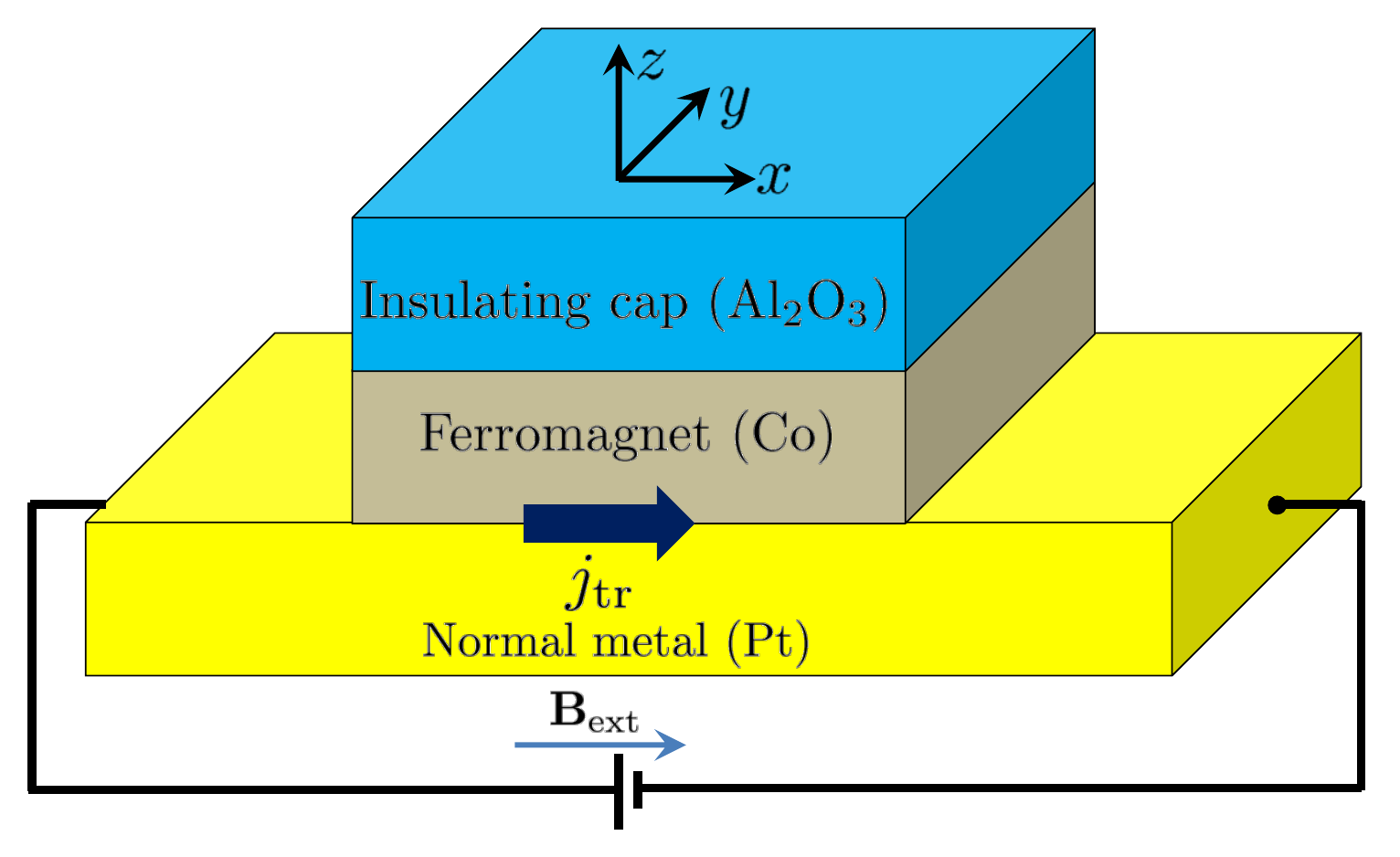}
\caption{(color online) Schematic view of trilayer devices for electrical magnetic switching.  A perpendicular-anisotropy
ferromagnetic layer (Co) is sandwiched between a metallic layer (for example Pt) and an insulating (
for example Al$_2$O$_3$) capping layer. The transport current, $j_{\rm{tr}}$, flows in the itinerant ferromagnet and normal metal in the direction parallel or antiparallel to the external magnetic field, $B_{\rm{ext}}$. }
\label{fig:setup}
\end{center}
\end{figure}

\noindent
{\em Introduction}---The central goal of spintronics is the discovery of efficient mechanisms for electrical
control of nanomagnet orientation.  The past fifteen years have witnessed many
advances (see reviews~\cite{RalphStiles,JungwirthRMP,Gambardella}, and references therein) that are potentially important for information storage technologies
and, because they depend on
improved understanding of non-equilibrium collective properties, also
interesting from a fundamental point of view.
A breakthrough has recently been achieved with the
demonstration~\cite{Miron2011,Ralph2011} of reliable electrically-controlled
magnetization switching in trilayer thin-film devices, illustrated schematically in Fig.~\ref{fig:setup},
that have a single perpendicular-anisotropy magnetic layer.
(Magnetization control with weak current induced torques had been observed previously\cite{Chernyshov2009} in (Ga,Mn)As.)
In the demonstration devices the ferromagnetic layer was Co, the normal metal layer Pt,
and the insulator Al$_2$O$_3$.  (Ref.~\onlinecite{Ralph2011} also studied other
insulators to partially demonstrate the universality of the effect.)


Although they appear to be observing the same effect, Refs.~\cite{Miron2011} and~\cite{Ralph2011} ascribe the current induced torques to different mechanisms.
Ref.~\cite{Miron2011} speculates that the effect is due to Rashba spin-orbit interactions of carriers in the ferromagnet and that it is analogous to the surprising semiconductor two-dimensional electron gases (2DEGs) $\hat{z}$ direction spin-densities induced by current flow parallel to an applied magnetic field~\cite{Kato,Engel2007}.  On the other hand, Ref.~\cite{Ralph2011} argued that the torques are due to a large spin Hall
current in Pt that flows vertically into the magnetic layer. The two effects are identical from the symmetry point of view, as already observed in Ref.~\cite{Miron2011}.

The spin-Hall-effect point of view has been elaborated on rather extensively in Ref.~\cite{Ralph2011}. The model we study in this paper is therefore most closely related to the Rashba effect interpretation. In general, current-induced torques in ferromagnets have for several years been recognized theoretically as a general consequence of spin-orbit coupling (see Refs.~\cite{ZhangPRB79_094422,Garate2009,NunezSSC2006}, as well as review~\cite{Gambardella}, and references therein).
For Rashba ferromagnets with $\hat{z}$ direction inversion symmetry breaking, out-of plane
spin-densities can be explained~\cite{Engel2007}  only by a full solution of the two-band
quantum kinetic equation.  With this motivation we report on a full quantum kinetic theory of current-induced torques in a ferromagnetic conductor with structural inversion asymmetry.

\noindent
{\em Rashba Ferromagnets}---
Quantitative microscopic theories of transport in transition metal ferromagnets
are complicated by the presence at the Fermi energy of several $3d$- and $4s$-derived electronic bands
that have complicated Fermi surfaces.
Most transport properties of collinear ferromagnetic conductors, including the important
giant and tunnel magnetoresistance effects, can nevertheless be understood qualitatively in terms of a
simple two-channel model\cite{Fert1,Fert2,Fert3,Fert4}
in which $\uparrow$-spin and $\downarrow$-spin carriers conduct in parallel with
conductances that depend strongly on the magnetization direction and its space dependence.
The direction dependence is due to a combination of spin-split
band structure and spin-dependent scattering effects.
We will see that the spin-dependent transport is also important for the present problem. We take the point of view that qualitative insights can be achieved by studying a simplified model that accounts for the interplay between structural inversion asymmetry and spin-dependent transport.

Cobalt is a strong ferromagnet~\cite{Wohlfarth} in which the majority-spin Fermi surface intersects only
a high-velocity free-electron-like ($s$) band with a substantial conductance, while the minority-spin Fermi surface intersects
several complex bands with generally small velocities, large densities-of-states, and dominant atomic-like d-character.
Minority-spin states at the Fermi energy have strongly hybridized free-electron-like and $d$ amplitudes.
Even high-velocity minority-spin states contribute less effectively to transport because they
scatter rapidly into the many available $d$-states.
To capture the essential physics we employ a phenomenological two-band model with short-lifetime minority spins and long-lifetime majority spins, neglecting the $d$-orbital contribution to transport as
is customary in theories of GMR~\cite{Fert1,Fert2,Fert3,Fert4}.
Scattering is spin-dependent because any crystal defect will produce different potentials for majority and minority electrons when strong exchange is taken into account, and because minority s-electrons can scatter into empty minority d-electron states. As is common in model studies, we add spin-orbit coupling and vertical structural inversion asymmetry to the two-band model by adding a Rashba SO term to the Hamiltonian.

These considerations lead to a Rashba-model ferromagnet  with band Hamiltonian:
\begin{eqnarray}\label{eq:s-hamiltonian}
  H_{\rm{s}}=\int dr \Psi^\dagger_\s\left(\hat{\epsilon}_\pp-\frac12\BB_\pp\ss+ \hat{U}_{\rm{dis}}\right)\Psi_\s.
\end{eqnarray}
Here $\hat{\epsilon}_\pp$ denotes the operator for the band energy of spin-orbit and exchange uncoupled itinerant electrons. Because the Co layer is thick enough ($\sim 10${\AA}) to support several 2D subbands,
we take $\pp$ to be a 3D vector.
The momentum-dependent effective magnetic field includes an exchange contribution in the direction $\mm$ of the
$d$-electron magnetization and a spin-orbit field in the $\hat{z} \times \pp$ direction:
$\BB_\pp=-\Delta_{\rm{xc}}\mm+2\alpha \, \hat{z}\times\pp$, $\alpha$ being the strength of the Rashba coupling.
(We neglect Zeeman coupling in comparison, assuming that it is important
only in helping to fix the direction of $\hat{m}$.)
To avoid confusion in what follows, we define $\mm$ to be the direction of local \emph{magnetization}. Thus the majority (minority) electrons have their spins antialigned (aligned) with the local magnetization,
which corresponds to $\Delta_{\rm{xc}}>0$.  $\hat{U}_{\rm{dis}}$ is the disorder potential. We adopt a short-range-disorder model with a
spin-dependence that is dictated by $\mm$:
\begin{equation}
  \hat U_{\rm{dis}}=\sum_i \delta(\rr-\rr_i)(u_\downarrow P^0_{+} + u_\uparrow P^0_{-}),
\end{equation}
where the index $i$ labels the impurity positions $\rr_i$,
$P^0_{\pm}=(1\pm \ss \cdot \mm)/2$ are projectors onto the local direction of the exchange field, and $u_{\uparrow}$ and $u_{\downarrow}$
characterize the strengths of majority and minority spin scattering, respectively.
Because majority-spin electrons have a higher mobility we must chose $u_\downarrow>u_\uparrow$.
Our model neglects inelastic phonon and magnon scattering which should be
relatively unimportant in highly disordered ultrathin films
even at room temperature. Below we refer to the high-velocity electrons described by the Rashba ferromagnet
model as the transport electrons, and to the spin-density of the $d$-electrons as the magnetization.

\noindent
{\em Current Induced Torques}---
Because the dynamics of the itinerant electrons in magnetic conductors is much faster than that of the magnetization,
it is sufficient to consider the kinetics of transport electrons in a static exchange field.

The effective magnetic field exerted on the magnetization by the
transport electrons is due to their mutual exchange interaction\cite{Garate2009} and therefore proportional to
$\Delta_{{\rm xc}}$:
\begin{equation}\label{eq:Heff}
  \HH_{\rm{eff}}=-\frac{1}{M_S}\frac{\delta \bra H\ket_{\rm{s}}}{\delta\mm}=-\frac{\Delta_{\rm{xc}}}{M_S}\, \bra\hat {\textbf{s}}\ket_{\rm{s}}.
\end{equation}
The subscript `$\rm{s}$'  on the angle brackets indicates the expected value in the transport steady state.
Here $\hat {\textbf{s}}$ is the transport electron spin-density operator
$ \hat{{\bf{s}}}= \hbar \Psi^\dagger_\sigma\ss_{\sigma\sigma'} \Psi_{\sigma'}/2$
and $M_{S}$ is the $d$-electron spin-density.  The theory of
current induced torques is therefore equivalent to a theory of $\bra\hat {\textbf{s}}\ket_{\rm{s}}$. The required transport theory differs essentially from the 2DEG problem considered in Ref.~\cite{Engel2007}, since the spin dependence of transport comes primarily from the spin-dependence of impurity scattering in an itinerant ferromagnet and the presence of a minority Fermi surface for $d$-like electronic bands, rather than solely from band-structure spin-splitting and angular dependence of scattering as in a 2DEG~\cite{Engel2007}. (Below we will also present the leading-order results for the latter case.) The component of $\bra\hat {\textbf{s}}\ket_{\rm{s}}$ perpendicular to the magnetization, which is responsible for magnetization switching in the devices of interest, depends essentially on current-induced interband coherence and
must be evaluated using a two-band quantum kinetic theory.

The kinetic equation can be obtained along the route outlined in Ref.\cite{Shytov2006}.  In this paper we do not, however, switch to the energy representation, choosing to work in the more intuitive $\pp$-representation.  The final form of the equations
we use is therefore slightly different.

To facilitate the discussion that follows, we first establish some notation. In the absence of disorder, the eigenvalues of Hamiltonian~(\ref{eq:s-hamiltonian}) are given by
\begin{equation}
  \epsilon_{\pp\nu}=\epsilon_\pp-\frac{1}{2}\nu B_\pp,
\end{equation}
where the index $\nu=\pm$ distinquishes majority ($+$)
and minority ($-$) bands.
The projection operators onto the corresponding eigenstates are
\begin{equation}
  P_{\pp\nu}=\frac{1}{2}\left(1+\nu\frac{\ss \cdot \BB_\pp}{B_\pp}\right).
\end{equation}

For weak disorder of the form specified in Eq.~(\ref{eq:s-hamiltonian}), the disorder self-energy is
\begin{equation}
  \check{\Sigma}(\rr)=n_i\hat\gamma \check G(\rr,\rr)\hat\gamma,\quad \hat\gamma=u_\downarrow P^0_{+} + u_\uparrow P^0_{-},
\end{equation}
where the {\em check} accent denotes matrices in Keldysh space~\cite{LL10}.

The quantum kinetic equation for the two-band density matrix $\hat{f}_\pp$,
linearized with respect to a time-independent uniform external electric field $\EE$, is
\begin{widetext}
\begin{equation}\label{eq:kineq}
  \partial_t \f_\pp+\frac{1}{2}\left\{\partial_\pp\epsilon_\pp -\frac{1}{2}\partial_\pp\BB_\pp\ss,\partial_\rr\f_\pp\right\} -\frac{i}{2}[\BB_\pp\ss,\f_\pp] +e\EE\partial_\pp\f^{\textrm{eq}}_\pp=\hat{I}_{st}.
 \end{equation}
Here  $\hat{f}_\pp$ is $2\times2$ matrix in spin-space and
$\f^{\textrm{eq}}_\pp=\sum_{\nu}P_{\pp\nu} f_{th}(\epsilon_{\pp\nu})$,
where $f_{th}(E)$ is the Fermi function, is the equilibrium distribution function.
Disorder effects appear in the collision integral $\hat{I}_{st}$ which
takes the form
 \begin{equation}\label{eq:collision}
  \hat{I}_{st}=-\pi n_i \sum_{\nu,\nu'}\int\frac{d^3p'}{(2\pi)^3} \delta(\epsilon_{\pp\nu}-\epsilon_{\pp'\nu'})
  \left(
  P_{\pp\nu}\f_\pp\hat\gamma P_{\pp'\nu'} \hat\gamma+
  \hat\gamma P_{\pp'\nu'} \hat\gamma \f_\pp P_{\pp\nu}-
  P_{\pp\nu}\hat\gamma \f_{\pp'} P_{\pp'\nu'} \hat\gamma-
  \hat\gamma P_{\pp'\nu'} \f_{\pp'}\hat\gamma P_{\pp\nu}
  \right).\nonumber
\end{equation}
\end{widetext}
It is easy to check that the collision integral vanishes when $\hat{f}_\pp$ is replaced by $\hat{f}^{\textrm{eq}}_\pp$.

In what follows, we assume that the spin splitting of the transport electrons $B_\pp$ is small compared to their
Fermi energy $E_F$.  In the analogous semiconductor 2DEG kinetic equation\cite{Engel2007} one has to keep at least the terms linear in $B_\pp/E_F$ in the kinetic equation in order to consistently describe the coupling between charge currents
and spin densities.  In the present case, however, the spin and charge densities are already
coupled at $(B_\pp/E_F)^0$ level because of spin-dependent scattering.
In order to isolate the essential physics as transparently as possible we initially neglect Fermi surface splitting
in the field-generation and collision terms.
It is legal to keep the precession term, which is linear in $B_\pp$, since it competes with relaxation rates only.
We comment further on the influence of spin-split bands below.

We now specialize to the stationary and uniform case.
After straightforward manipulations, using $\sum_\nu P_{\pp\nu}=1$, performing the integrals over $\pp'$ in the collision integral, and representing the distribution function as a sum of scalar and vector parts using
 $\f_\pp=n_\pp+\ss\cdot\ff_\pp$, we obtain the following
equations:
\begin{eqnarray}\label{eq:kineq_SDS}
e\EE\partial_\pp f_{th}&=&-\frac{1}{\tau_\textrm{s}}(n_\pp-\overline {n}_{\pp})-\frac{\mm}{\tau_\textrm{d}} \cdot (\ff_\pp-\overline\ff_{\pp}),\nonumber\\
B_\pp\times\ff_\pp&=&-\frac{\mm}{\tau_\textrm{d}} \, (n_\pp-\overline n_{\pp})-\frac{1}{\tau_\textrm{s}}(\ff_\pp-\overline \ff_{\pp})\nonumber\\
&&-\frac{1}{\tau_\perp}\mm\times\overline \ff_{\pp}\times \mm,
\end{eqnarray}
where $1/\tau_{\rm{s,d}}=\pi n_iN_0(u_\downarrow^2 \pm u_\uparrow^2)>0$, $N_0$ being the density of states at the Fermi level, and the overbar accent
denotes the average over the directions of $\pp$.
The spin-decoherence rate, $1/\tau_\perp$,  {\em i. e.} the rate of local relaxation of spin-components
perpendicular to $\mm$,  is equal to $\pi n_iN_0(u_\downarrow-u_\uparrow)^2$ in this model, and thus is a function of $1/\tau_{\rm{s,d}}$. We keep it as an independent parameter to recognize that this property does not hold for general spin-dependent disorder models~\footnote{Since Pt atoms are effective spin scatterers, and Co and Pt are quite miscible, we should have included  Elliot-Yafet-type of spin-orbit scattering by impurities. This would contribute to the spin decoherence rate,  as well as introduce relaxation rate for the diagonal component of the local in space spin polarization (L. Berger, Phys. Rev. B 83, 054410 (2011)). However, relaxation of non-zero angular harmonics of the distribution function would be still dominated by the terms already present in equation~(\ref{eq:kineq_SDS}). Thus we would like to keep $\tau_\perp$ as an independent parameter, instead of complicating the kinetic equation unnecessarily.}. Longitudinal spin density (aligned with $\mm$) is not generated to linear order in the spin-orbit interaction in the present case and we therefore do not need to the introduce the corresponding relaxation time.

The solution of Eq.~(\ref{eq:kineq_SDS}) is straightforward.  The scalar equation is used to
eliminate the charge response from the vector equation, thereby
introducing a generation term and an additional relaxation term in the equation for $\ff_\pp$.
To linear order in $\alpha$, we first obtain an equation for the longitudinal spin response
(the component of $\ff_\pp$ parallel to $\mm$), which does not directly produce a torque.
When this solution is substituted into the equation for the transverse response
it adds a generation term to the equation for the perpendicular component because of its precession in the spin-orbit field.
The current-induced spin-densities and currents are then obtained by summing the scalar and vector
responses over $\pp$.

Using Eq.~(\ref{eq:Heff}), we finally obtain the following expressions for the effective magnetic fields that act on the ferromagnet's magnetization, and for the current density flowing the in the film:
\begin{eqnarray}\label{eq:fieldFM}
   \HH^{\rm{R}}_{\textrm{eff}}&=&\frac{\alpha mj_{{\rm{tr}}}}{|e| M_S} \frac{\tau_\uparrow-\tau_\downarrow}{\tau_\uparrow+\tau_\downarrow} \frac{\Delta_{\textrm{xc}}^2\tau_\perp^2} {1+\Delta_{\textrm{xc}}^2\tau^2_{\perp}} \mm\times\hat{y}\times\mm,\nonumber\\
  \HH^{\rm{S}}_{\textrm{eff}}&=&\frac{\alpha mj_{{\rm{tr}}}}{|e| M_S} \frac{\tau_\uparrow-\tau_\downarrow}{\tau_\uparrow+\tau_\downarrow} \frac{\Delta_{\textrm{xc}}\tau_\perp} {1+\Delta_{\textrm{xc}}^2\tau^2_{\perp}} \hat{y}\times\mm,\nonumber\\
  \textbf{j}_{\rm{tr}}&=&\frac{n_{\rm{tr}}e^2}{m} (\tau_\uparrow+\tau_\downarrow)\textbf{E}.
\end{eqnarray}
where $j_{\rm{tr}}$ is the transport current density, $n_{\rm{tr}}$ is the transport electron density per spin; we assumed for simplicity a parabolic spectrum of transport electrons, and introduced $m$, the correpsonding band mass (the same for majority and minority electrons). The latter assumption is not crucial, and has been invoked only for reasons of clarity of the expressions. Finally, $\tau_{\downarrow,\uparrow}^{-1} = \tau_{\rm{s}}^{-1} \pm \tau_{\rm{d}}^{-1}$
are the minority ($1/\tau_{\downarrow}$) and  majority ($1/\tau_{\uparrow}$) spin scattering rates. These expression are the main result of this work.
We identify the field $\HH^{\rm{R}}_{\textrm{eff}}$, which produces a torque identical to a field in the $\hat{y}$ direction and
cannot switch a perpendicular film as the Rashba field.
We note that the expression for the Rashba field given here is parametrically larger than the one in existing
literature~\cite{ZhangPRB79_094422}, which contains an additional $\Delta_{\rm{xc}}/E_F$ smallness factor. If Ref.~\cite{Miron2010} was analyzed using our expression it would
decrease the experimental estimate of the Rashba coupling parameter.
The field $\HH^{\rm{S}}_{\textrm{eff}}$ has the same dependence on magnetization as the effective field produced by a $\hat{y}$-polarized spin-current flowing into the magnetic layer, and can switch perpendicular currents~\cite{Miron2011,Ralph2011}.

In the general case of spin-dependent scattering and exchange fields of arbitrary strength the two-band quantum kinetic equation cannot be solved analytically.  There are however other limits in which instructive analytic results can be obtained. One interesting limit is that of a 2D Rashba ferromagnet with spin-independent disorder and a spin-splitting that is larger than the Bloch state lifetime. Results for the effective field, ${\bf{h}}_{\textrm{eff}}$,  to the leading order in $\alpha$ and in the clean limit, $B_\pp\tau\gg 1$, are
\begin{eqnarray}\label{eq:field2DEG}
  {\bf{h}}^{\rm{R}}_{\textrm{eff}}&=&\frac{\alpha m j_{\textrm{tr}}}{2|e|M_S}\frac{\Delta_{\textrm{xc}}}{E_F}\hat{y},\nonumber\\
  {\bf{h}}^{\rm{S}}_{\textrm{eff}}&=&-\gamma_0\frac{\alpha m j_{\textrm{tr}}}{2|e|M_S}\frac{\Delta_{\textrm{xc}}}{E_F} \frac{1}{\Delta_{\textrm{xc}}\tau}\frac{m_xm_z}{1+m_z^2}\mm.
\end{eqnarray}
In the above expressions $\gamma_0=\left(\frac{p_F}{v_F}\pder{v_F}{p_F}-1\right)$ is the nonparabolicity parameter, introduced in Ref.~\cite{Engel2007}. In Eqs.~(\ref{eq:field2DEG}), the field ${\bf{h}}^{\rm{R}}_{\textrm{eff}}$ is the Rashba field found earlier in Ref.~\cite{ZhangPRB79_094422}. The field ${\bf{h}}^{\rm{S}}_{\textrm{eff}}$, despite being aligned with the magnetization, can in principle lead to switching since it is odd in $m_z$. The factor $1+m_z^2$ in the denominator of the expression for ${\bf{h}}^{\rm{S}}_{\textrm{eff}}$ can be traced to the fact that for a 2DEG with a Rashba SO, the Dyakonov-Perel~\cite{DyakonovPerel} in-plane spin-relaxation time is twice as long as the out-of-plane one. We have dropped contributions to ${\bf{h}}_{\textrm{eff}}$ that are proportional to $\gamma_0$ but even in $m_z$ since they cannot lead to switching. Terms with the symmetry of ${\bf{H}}^{\rm{S}}_{\textrm{eff}}$ from Eq.~(\ref{eq:fieldFM}) do appear at $O(\alpha^3)$ order, and are also proportional to $\gamma_0$, but the corresponding expressions are cumbersome and are not shown.

By examining these two analytically accessible limits, we have concluded that the crucial ingredient in the switching behavior is spin-dependent scattering.  It seems likely to us that the most important source of this spin-dependence in Co is the presence at the Fermi energy of minority spin d electrons which is captured in Eqs.~\ref{eq:fieldFM}.

\noindent
\textit{Discussion - } The physical origin of the effects described here is the fact that when an electric field is applied to a ferromagnet with different majority and minority mobilities, the two spin species react to it differently. This creates spin polarization in momentum space, whose angular dependence ``matches'' the angular dependence of the y component of Rashba SO field. Thus a component of spin polarization perpendicular to the ferromagnet's magnetization is generated. This component then precesses around the exchange field, and gets relaxed either by magnetic scattering or the Dyakonov-Perel mechanism. A similar scenario was outlined previously~\cite{Engel2007} in connection with the explanation for out-of-plane spin polarizations in in-plane fields in 2DEGs with Rashba interactions.

Despite the similar underlying mechanism, the leading order results in the two models we considered, Eqs.~(\ref{eq:fieldFM}) and~(\ref{eq:field2DEG}) have different form. In particular, Eqs.~(\ref{eq:fieldFM}) correspond to the case of spin-dependent impurity scattering, and  Eqs.~(\ref{eq:field2DEG}) are those for the case of a spin-split band structure and spin-independent impurity scattering. The difference is thus not surprising: since the origin of the spin-dependent scattering is different in the two models, generation and relaxation terms have different dependence on the Rashba SO strength.

There is some correspondence between the current-induced fields and torques that we have evaluated for a Rashba ferromagnet model,  and those found experimentally. The torques exerted on the magnetization by $\HH^{\rm{S}}_{\rm{eff}}$, Eqs.~(\ref{eq:fieldFM}), have the same symmetry properties as the experimental current-induced torques in Refs.\cite{Miron2011} and~\cite{Ralph2011}. In addition, the field has the same sign as observed experimentally,
since it implies switching from up to down direction for current flowing in the direction of the applied external in-plain magnetic field, if $\alpha$ is positive. The fact that it is positive in the experiments of Miron \textit{et al.} follows from their observation that $\HH^{{\rm{R}}}_{\rm{eff}}$ points in the positive $\hat{y}$ direction (the coordinate system used in this work is identical to theirs).
This universality is ensured by the transport properties of Co, in which the majority electrons have a much larger lifetime. In the case of a 2DEG, Eqs.~(\ref{eq:field2DEG}), the direction of the switching field is less universal, in the sense that it depends on the details of the band structure through the nonparabolicity parameter $\gamma_0$.

The magnitudes of the effective fields in a Rashba ferromagnet, Eqs.~(\ref{eq:fieldFM}), depend on two  parameters, $\alpha$ and $\Delta_{\rm{xc}}\tau_\perp$, which could be estimated by detailed first principles studies of Co/Pt multilayers in the presence of a structural asymmetry. By comparing the ratio of switching and Rashba fields in Eqs.~(\ref{eq:fieldFM}), we observe that the values of $\HH^{\rm{R}}_{\rm{eff}}/j_{{\rm{tr}}}\approx 10^{-8}{{\rm{T cm^2A^{-1}}}}$ claimed in Ref.~\cite{Miron2010}, and $\HH^{\rm{S}}_{\rm{eff}}/j_{{\rm{tr}}}\approx 10^{-9}{{\rm{T cm^2A^{-1}}}}$, claimed in Ref.~\cite{Miron2011} in a similar setup, yield a plausible value of $\Delta_{{\rm{xc}}}\tau_\perp\sim 10$.

For a large $\Delta_{\textrm{xc}} \tau_{\perp}$ our results for the
Rashba spin-density correspond to
\begin{equation}
\frac{s^{{\rm{R}}}}{n_{{\rm{tr}}}} \sim \frac{\alpha p_F}{\Delta_{{\rm{xc}}} } \; \frac{v_{D}}{v_{F}}
\end{equation}
where $v_{D} \sim e E \tau /m$ is the transport drift velocity.  In Co we estimate that
$v_{D} \sim 100\, {\rm{m/s}}$ at a current density of $10^{12} {\rm{A/m^2}}$, compared to a
transport electron Fermi velocity $\sim 10^{6}\, {{\rm{m/s}}}$.  To estimate that Rashba coupling parameter of
Co on Pt we note that the majority-spin Fermi level is $\sim 0.5 eV$ above the top of the $d$-bands in Co and
and below the top of the band in Pt.  This implies an electric potential drop comparable to the Fermi energy over the thickness
$t_{{\rm{Co}}}$ of the Co layer and a Rashba coupling strength at the Fermi energy $\alpha p_{F}$ that is
$ \sim E_F^2/(mc^2  p_{F}t_{{\rm{Co}}})$ which should be around $10^{5}$ times smaller than
$ \Delta_{\textrm{xc}}$.   We conclude that it is hard to explain dimensionless current-induced spin-densities much larger than $\sim 10^{-9}$ at $10^{12} {\rm{A/m^2}}$ in Co based on this bulk mechanism.
On the other hand the quantitative analysis in Ref.\cite{Ralph2011} suggests a switching dimensionless spin-density at this current level that is $\sim 10^{-6}$. We thus note that a substantial enhancement of SO interactions is needed to explain the experimental switching current densities based on the Co Rashba mechanism. This enhancement could come from the hybridization of Co transport electrons  with Pt d-orbitals, analogous to SO enhancement by surface alloying with heavy elements~\cite{Ast}. This hybridization must play an essential role in the switching effect in both Rashba coupling and spin-Hall scenarios.  We therefore propose that the effect is strongest when these bands are closely aligned.  If so, this may explain the efficacy of Co/Pt structures and also suggest strategies for optimizing the effect.

The Authors acknowledge stimulating discussions with G.E.W. Bauer.
This work was supported by Welch Foundation
grant TBF1473 and by the ARO MURI on bioassembled nanoparticle arrays.

\end{document}